\documentclass[a4paper,11pt]{article}
\pdfoutput=1 % if your are submitting a pdflatex (i.e. if you have
\usepackage{jheppub} % for details on the use of the package, please
\usepackage{empheq}
\usepackage{bm}
\usepackage{slashed}
\usepackage[T1]{fontenc} % if needed
\usepackage[all]{xy}
\usepackage{rotating}
\usepackage{amsmath} 
\usepackage{mathtools}

\usepackage{slashed}

\def\CA{{\cal A}}

\def\CI{{\cal I}}

\def\CN{{\cal N}}

\def\CR{{\cal R}}
\def\CS{{\cal S}}
\def\CT{{\cal T}}

\def\beq#1\eeq{\begin{align}#1\end{align}}

\title{\boldmath Magnetically charged $AdS_5$ black holes from class $\CS$ theories on hyperbolic 3-manifolds}

\author[a]{Jin-Beom Bae,}
\author[b]{Dongmin Gang,}
\author[a]{and Kimyeong Lee}

\affiliation[a]{School of Physics, Korea Institute for Advanced Study, Seoul 02445, Republic of Korea}
\affiliation[b]{Quantum Universe Center, Korea Institute for Advanced Study, Seoul 02455, Korea}

\abstract{We study the twisted index of 4d $\mathcal{N}=2$ class $\CS$ theories on a closed hyperbolic 3-manifold $M_3$. %The index can be written in terms of simple topological invariants, analytic torsions twisted by irreducible flat connections, on the 3-manifold.
Via 6d picture, the index can be written in terms of  topological invariants called analytic torsions twisted by irreducible flat connections on the 3-manifold.  Using the topological expression, we determine  the full perturbative $1/N$ expansion of the twisted index. The leading part nicely matches the Bekestein-Hawking entropy of a magnetically charged black hole in the holographic dual $AdS_5$ with $AdS_2 \times M_3$ near-horizon.}

\begin{document} 
\begin{flushright}
{\tt\normalsize KIAS-P19038}\\
\end{flushright}
\maketitle

\section{Introduction and Conclusion}

Microscopic understanding of black hole entropy is one of the prominent success of string theory. Indeed, it is well known that string theory can provide microscopic interpretation to the Bekenstein-Hawking entropy of asymptotically flat black hole \cite{Strominger:1996sh}.
A lot of work has been done \cite{Dijkgraaf:1996it, Shih:2005uc, David:2006yn} in order to analyze black hole entropy involving quantum corrections, based on 2d field theory approach. Meanwhile, the entropy of black hole in AdS$_3$ has been analyzed via AdS/CFT correspondence \cite{Maldacena:1997re}, by counting microscopic states of 2d conformal field theory(CFT) \cite{Kraus:2006nb}.

It has been believed that entropy of higher-dimensional supersymmetric black hole in AdS$_d$($d>3$) can be  understood from  boundary superconformal field theory(SCFT) using AdS/CFT. Recently, there has been remarkable progresses in this direction. In \cite{Benini:2015noa, Benini:2015eyy}, the entropy of static dyonic BPS black hole in AdS$_4 \times S^7$ is shown to agree with the topologically twisted index of 3d ABJM model \cite{Aharony:2008ug} with $k=1$. More recently, many works have been done regarding entropy of black holes in AdS$_5$ \cite{Cabo-Bizet:2018ehj,Choi:2018hmj,Choi:2018vbz, Benini:2018ywd, Honda:2019cio, ArabiArdehali:2019tdm, Kim:2019yrz, Cabo-Bizet:2019osg} using 
refined 4d superconformal index.

%More recently, the authors of \cite{Cabo-Bizet:2018ehj,Choi:2018hmj} studied a large $N$ limit of  superconformal index for 4d $\mathcal{N}=4$ Yang-Mills theory to capture entropy of electrically charged black hole in AdS$_5\times S^5$. They tuned the imaginary part of chemical potential to hinder the cancellation between boson and fermion, then it turns out that the  index correctly reproduce the Bekenstein-Hawking entropy of the AdS$_5$ black hole. 
%Since then, many works have been done regarding entropy of black holes in AdS$_5$ \cite{ Choi:2018vbz, Benini:2018ywd, Honda:2019cio, ArabiArdehali:2019tdm, Kim:2019yrz, Cabo-Bizet:2019osg} using 4d superconformal index.

In this paper, our goal is to understand microscopic origin of a magnetically charged black hole \cite{Nieder:2000kc} in AdS$_5$   using twisted index of 4d $\mathcal{N}=2$ SCFT on $S^1 \times M_3$, where $M_3$ is a closed hyperbolic 3-manifold. 
The entropy of magnetically charged black holes of our interest is not easy to analyze quantitatively via localization technique, as we consider the 4d SCFT on closed hyperbolic 3-manifolds. To circumvent this technical difficulty, we suggest alternative way of computing twisted index of a certain class of 4d $\mathcal{N}=2$ SCFTs on $S^1 \times M_3$. We start at 6d  $(2,0)$ theory on $S^1 \times M_3 \times \Sigma_g$, where $\Sigma_g$ denote Riemann surface with genus $g$. As we shrinking the size of $\Sigma_g$, the 6d mother theory is reduced to 4d class $\CS$  SCFT   \cite{Gaiotto:2008cd,Gaiotto:2009we} on $S^1 \times M_3$. On the other hand, one can arrive to a 3d $\mathcal{N}=2$ class $\CR$ SCFT \cite{Terashima:2011qi,Dimofte:2011ju} on $S^1 \times \Sigma_g$ by reducing the size of $M_3$ in 6d theory. Since the 6d twisted index is invariant under continous supersymmetry preserving deformations, we expect  the equality between the 4d twisted index for class $\CS$ theory and 3d twisted index for class $\CR$ theory. Fortunately, the 3d twisted index
 was already studied  in \cite{Gang:2018hjd,Gang:2019uay} using the 3d-3d relation. Thus our main claim in this paper is that one can utilize the twisted index computation to analyze the entropy of magnetically charged black hole in AdS$_5$. At large-$N$ limit, we checked our speculation using supergravity solution in \cite{Nieder:2000kc,Maldacena:2000mw}.

One interesting future problem is to widen our understanding of the higher-dimensional black hole entropy up to subleading order in the $1/N$ expansion. We believe that the magnetically charged black hole in AdS$_5$ is an excellent testground for it, because subleading correction of twisted index already computed in \cite{Gang:2018hjd,Gang:2019uay}. It would be interesting to understand the meaning of those subleading correction at the M-theory side.

\section{Class $\CS$ theories on hyperbolic 3-manifold}
\subsection{Twisted index of class $\CS$ theories on  3-manifold $M_3$}
A 4d  class $\mathcal{S}$ theory $\CT_N[\Sigma_g]$ associated to a compact Riemann surface $\Sigma_g$ of genus $g$ is defined as \cite{Gaiotto:2008cd,Gaiotto:2009we}
\begin{align}
\begin{split}
\CT_N [\Sigma_g] &:= (\textrm{4d $\mathcal{N}=2$ SCFT at the infra-red (IR) fixed point of  } 
\\
&\qquad  \textrm{a twisted compactification of 6d $A_{N-1}$ (2,0) theory along $\Sigma_g$})\;. 
\end{split}
\end{align}
Using $SO(2)$ subgroup of $SO(5)$ R-symmetry of the 6d theory, we perform a partial twisting along $\Sigma_g$. It means that the following background gauge fields $A^{\rm background}_{SO(2)}$ coupled to the $SO(2)$ R-symmetry are turned on 
\begin{align}
\textrm{topological twisting : }A^{\rm background}_{SO(2)} = \omega (\Sigma_g),
\end{align}
where $\omega (\Sigma_g)$ is the spin connection on the Riemann surface. 
The topological twisting preserves 8 supercharges (16 supercharges) out of original 16 supercharges for $g > 1$ ($g=1$). For $g\geq 1$, the system is believed to flow to a non-trivial SCFT under the renormalization group. Especially when $g>1$, the $SO(5)$ R-symmetry of 6d theory is broken to $SO(2)\times SO(3)$ due to the topological twisting. The remaining symmetry can be identified with $u(1)_R \times su(2)_R$ R-symmetry of the 4d $\CT_N[\Sigma_g]$ theory.
\begin{align}
\begin{split}
&SO(5) \textrm{ R-symmetry of 6d $A_{N-1}$  (2,0) theory }
\\
& \xrightarrow{\quad \textrm{twisted compactification} \quad } SO(2)\times SO(3) \textrm{ R-symmetry of 4d $\CT_N[\Sigma_g]$ theory }
\end{split} 
\end{align}
 For sufficiently large $N$, the system does not have any emergent IR symmetry and the 4d $\CT_N[\Sigma_{g> 1}]$ is a $\CN=2$ SCFT without any (non-R) flavor symmetry. Then, the superconformal $u(1)_R$ symmetry can be identified with the compact $SO(2) \subset SO(2)\times SO(3)$.\footnote{For small $N$, there could be  accidental IR symmetries in $\CT_{N}[\Sigma_g]$ theory. }  
 
 We consider following twisted index of the $\CT_N[\Sigma_g]$ theory that is defined on closed hyperbolic 3-manifold $M_3 = \mathbb{H}^3/\Gamma$.
\begin{align}
\begin{split}
\mathcal{I}_{M_3} (\CT_{N}[\Sigma_g]) :=& \mathcal{Z}_{BPS}(\textrm{4d $\CT_{N}[\Sigma_g]$ on $S^1\times M_3$})
\\
=& \textrm{Tr} (-1)^R\;
\end{split}
\end{align}
Here, the trace is taken over the Hilbert-space of $\CT_N[\Sigma_g]$ on $M_3$. $\mathcal{Z}_{BPS}(\textrm{$\CT$ on $\mathbb{B}$})$ denotes the  partition function of a theory $\CT$ on a supersymmetric background $\mathbb{B}$ while 
$R$ denote the charge of IR superconformal  $u(1)_R$ R-symmetry, which is normalized as
\begin{align}
R (Q) = \pm 1\;, \quad \textrm{for supercharge $Q$}\;.
\end{align}
Note that the topological twisting is performed along the $M_3$ using the $su(2)_R$ symmetry of the 4d SCFT, to preserve some supercharges.

The twisted index can be defined for arbitrary 4d $\CN=2$ SCFTs using $su(2)_R \times u(1)_R$ R-symmetry. For general 4d $\CN=2$ SCFTs, the charge $R$ is not integer valued and thus the index is %generally 
complex valued. For $\CT_N[\Sigma_g]$ theory with sufficiently large $N$, on the other hand, the twisted index is an integer because the $u(1)_R$ symmetry comes from compact $SO(5)$ R-symmetry of 6d theory. The index is invariant under the continuous supersymmetric deformations of the 4d theory, thus it only depends on the topology (not on the metric choice)  of $M_3$. 
\begin{align}
\mathcal{I}_{M_3} (\mathcal{T}_N[\Sigma_g]) \;:\; \textrm{a topological invariant of $M_3$}\;
\end{align}
In the next  section, we  express the twisted index $\CI_{M_3}(\mathcal{T}_N[\Sigma_g])$ in terms of  previously known topological invariants on 3-manifold, {\it analytic torsion} twisted by irreducible flat connections. 
%The non-existence of the flavor symmetry make story much simpler when comparing supersymmetric indices (superconformal index or twisted index) of $\CT_{N}[\Sigma_g]$ with microstates counting of corresponding black hole in the gravity dual. ({\bf DG : Edited until here })  

\subsection{Twisted Index computation using 6d picture  }
For a generic 4d $\CN=2$ SCFT, which we will denote as $\CT_{4d \;\CN=2}$, the computation of the twisted index $\CI_{M_3}(\CT_{4d\;\CN=2})$  on  a hyperbolic manifold $M_3$ is quite challenging. 
Unlike 3d cases \cite{Nekrasov:2014xaa,Benini:2015noa,Benini:2016hjo,Closset:2016arn,Closset:2017zgf}, there is no developed localization formula for the 4d twisted index on hyperbolic 3-manifolds.  Since the landscape of hyperbolic 3-manifolds is much wilder than hyperbolic Riemann surface, obtaining localization formula for general $M_3$ might be very challenging task. For class $\CS$ theories $\CT_N [\Sigma_g]$, we can bypass these  difficulties using 6d picture.  

\paragraph{A 4d-3d relation} The twisted index for 4d class $\mathcal{S}$ theory $\CT_N[\Sigma_g]$ on $M_3$ can be obtained from the supersymmetric partition function of 6d $A_{N-1}$ (2,0) theory on $S^1\times M_3 \times \Sigma_g$ by shrinking the size the $\Sigma_g$: 
\begin{align}
\begin{split}
&\mathcal{Z}_{BPS}(\textrm{6d $A_{N-1}$ (2,0) theory on $S^1\times M_3\times \Sigma_g$})
\\
& \xrightarrow{\quad \Sigma_g \rightarrow 0 \quad }  \mathcal{Z}_{BPS}(\textrm{4d $\CT_N[\Sigma_g]$ on $S^1\times M_3$})  \;.
\end{split}
\end{align}
As usual Witten index, the 6d partition function is invariant under the continuous supersymmetric deformations. One possible supersymmetric deformation is the overall size change of the compact Riemman surface $\Sigma_g$. Thus, we expect 
\begin{align}
\begin{split}
&\mathcal{Z}_{BPS}(\textrm{6d $A_{N-1}$ (2,0) theory on $S^1\times M_3\times \Sigma_g$})
\\
& = \mathcal{Z}_{BPS}(\textrm{4d $\CT_N[\Sigma_g]$ on $S^1\times M_3$})  \;. \label{6d-4d relation}
\end{split}
\end{align}
On the other hand, one may consider a limit where the size of $M_3$ shrinks. In the case, from the same argument above we expect
\begin{align}
\begin{split}
&\mathcal{Z}_{BPS}(\textrm{6d $A_{N-1}$ (2,0) theory on $S^1\times M_3\times \Sigma_g$})
\\
& = \mathcal{Z}_{BPS}(\textrm{3d $\CT_N[M_3]$ on $S^1\times \Sigma_g$})  \;. \label{6d-3d relation}
\end{split}
\end{align}
Here  $\CT_N[M_3]$ is a 3d  class $\mathcal{R}$ theory associated to a hyperbolic 3-manifold $M_3$:
\begin{align}
\begin{split}
\CT_N [M_3] &:= (\textrm{3d $\mathcal{N}=2$ SCFT at the IR fixed point of a twisted compactification } 
\\
&\qquad  \textrm{of 6d $A_{N-1}$ (2,0) theory along $M_3$})\;. 
\end{split} 
\end{align}
In the partial topological twisting along $M_3$, we use a $SO(3)$ subgroup of $SO(5)$ R-symmetry of the 6d theory and the twisting  preserves 1/4 supercharges. Therefore, the resulting low-energy theory is described by  a  3d $\mathcal{N}=2$  SCFT denoted as $\CT_N[M_3]$.
Combining  \eqref{6d-4d relation} and \eqref{6d-3d relation}, we  have a  following equality between the 4d/3d twisted indices
\begin{align}
\textrm{`4d-3d relation' : }\mathcal{I}_{M_3} ( \CT_N[\Sigma_g]) = \mathcal{I}_{\Sigma_g} (\CT_{N}[M_3])\;. \label{4d-3d relation}
\end{align}
Here, the 3d twisted index $\CI_{\Sigma_g} (\CT_N[M_3])$ is defined as
\begin{align}
\begin{split}
\mathcal{I}_{\Sigma_g} (\CT_{N}[M_3]) :=&  \mathcal{Z}_{BPS}(\textrm{3d $\CT_{N}[M]$ on $S^1\times \Sigma_g$})
\\
=&\textrm{Tr} (-1)^R\;, 
\end{split}
\end{align}
where the trace is taken over the Hilbert-space of $\CT_N[M_3]$ on $\Sigma_g$.

%In summary, we derive the 4d-3d relation  \eqref{4d-3d relation} from following diagram:
In summary, the 4d-3d relation  \eqref{4d-3d relation} is depicted in the diagram below :
\vspace*{1mm}
\begin{displaymath}
\boxed{
\xymatrix{
	\mathcal{Z}_{BPS}(\textrm{6d $A_{N-1}$ (2,0) theory on $S^1\times M_3\times \Sigma_g$})  \ar[d]^{M_3 \rightarrow 0} \ar[dr]^{\Sigma_g \rightarrow 0} & \\ 
	\mathcal{Z}_{BPS}(\textrm{3d $\CT_N[M_3]$ on $S^1\times \Sigma_g$})  &  \mathcal{Z}_{BPS}(\textrm{4d $\CT_N[\Sigma_g]$ on $S^1\times M_3$})
}
}
\end{displaymath}
\vspace*{0.1mm}
See \cite{Gukov:2016gkn} for previous study on  the relations among supersymmetric partition functions in various dimensions originated from the same 6d picture.

\paragraph{$\mathcal{I}_{\Sigma_g} ( \CT_N[M_3])$ from twisted analytic torsions on $M_3$} The quantity on the RHS of \eqref{4d-3d relation} is much easier to handle than the quantity on the LHS. First, we have an explicit field theoretic description for 3d theory $\CT_N[M_3]$ for general $N \geq 2$ and %general 
closed hyperbolic 3-manifold $M_3$ \cite{Dimofte:2011ju,Dimofte:2013iv,Gang:2018wek}. Second, we have general localization formula for the 3d twisted index.  Combining these  developments, recently it was found that the twisted index $\mathcal{I}_{\Sigma_g}(T_N[M_3])$ is  simply given as \cite{Gang:2018hjd,Gang:2019uay}:
\begin{align}
\mathcal{I}_{\Sigma_g} (\CT_{N}[M_3]) = \sum_{\CA^\alpha \in \chi^{\rm irred}(M_3;N)} (N  \times {\bf Tor}^{\alpha}_{M_3})^{g-1}\;. \label{index from torsion}
\end{align}
For a technical reason, the above relation holds only for closed hyperbolic 3-manifold with vanishing $H_1 (M_3, \mathbb{Z}_N)$ \cite{Gang:2019uay}.\footnote{Generalization to general $M_3$ is proposed in \cite{Benini:2019dyp}.
 }
The summation is over irreducible $PGL(N,\mathbb{C})$ flat connections on $M_3$:
\begin{align}
\chi^{\rm irred}(M_3;N) := \frac{\{ \textrm{irreducible $PGL(N,\mathbb{C})$ flat-connections on $M_3$}\}}{(\textrm{gauge quotient})}\;.
\end{align}
The set is  finite  for generic choice of $M_3$.
${\bf Tor}^{\alpha}_{M_3}$ is a topological invariant called analytic (or Ray-Singer) torsion twisted by a flat connection $\mathcal{A}^\alpha$. The topological quantity is defined as follows \cite{ray1971r,gukov2008sl}
\begin{align}
{\bf Tor}^\alpha_{M_3} := \frac{[\det ' \Delta_1 (\mathcal{A}^\alpha)]^{1/4}}{[\det ' \Delta_0 (\mathcal{A}^\alpha)]^{3/4}}\;.
\end{align}
where $\Delta_n (\mathcal{A}^\alpha)$ is a Laplacian acting on $pgl(N,\mathbb{C})$-valued $n$-form twisted by a flat connection  $\mathcal{A}^\alpha$:
\begin{align}
\Delta_n (\mathcal{A}) = d_A * d_A * +*d_A *d_A\;, \quad d_A  = d+ \mathcal{A}\wedge\;.
\end{align}
 One non-trivial  prediction  of the  above relation is that the topological quantity on RHS is an integer. The integrality has been checked for various examples in \cite{Gang:2019uay}, which gives a strong consistency check for the 3d-3d relation \eqref{index from torsion}.
 %Combining
 
 From \eqref{4d-3d relation} and \eqref{index from torsion}, we finally have following simple expression for the  twisted index of 4d class S theories $\CT_N[\Sigma_g]$ on closed hyperbolic 3-manifold $M_3$ with $H_1(M_3, \mathbb{Z}_N)=0$:
 \begin{align}
 \mathcal{I}_{M_3} ( \CT_N[\Sigma_g]) =  \sum_{\CA^\alpha \in \chi^{\rm irred}(M_3;N)}  (N  \times {\bf Tor}^{\alpha}_{M_3})^{g-1}\;.  \label{4d twisted index from torsions}
 \end{align}

\section{Large $N$ twisted index and a magnetically charged $AdS_5$ BH} 
Via AdS$_5$/CFT$_4$, the twisted index is expected to count the  microstates of  a magnetically charged black hole in the holographic dual AdS$_5$ gravity with {\it  signs}:
\begin{align}
\begin{split}
  \CI_{M_3}(\CT_N[\Sigma_g]) &=d^{+}_{\rm micro} - d^{-}_{\rm micro},
\end{split}
\end{align}
here $d^{+}_{\rm micro}$ and $d^{-}_{\rm micro}$ stand for the number of microstates with even $R$-charge and odd $R$-charge, respectively. 
Unless there is highly fine-tuned cancellation between $d^{+}_{\rm micro}$ and $d^{-}_{\rm micro}$, the twisted index can see the Bekenstein-Hawking entropy $S_{BH}$ of the black hole at large $N$:
\begin{align}
\begin{split}
&\log  \CI_{M_3}(\CT_N[\Sigma_g])  = \log  (d^{+}_{\rm micro}-d^{-}_{\rm micro}) \simeq \log (d^{+}_{\rm micro}+d^{-}_{\rm micro}) \simeq  S_{BH}, \;\;
\\
&\textrm{unless}\;\;  \bigg{|} \frac{d^{+}_{\rm micro}-d^{-}_{\rm micro}}{d^{+}_{\rm micro}+d^{-}_{\rm micro}} \bigg{|}  <  e^{ -\kappa N^3 }\; \textrm{at sufficiently large $N$ for some positive, finite $\kappa$}.
\end{split}
\end{align}
Here the equivalence relation $\simeq$ is defined as
\begin{align}
f(N)\simeq g(N) \;\; \textrm{if } \lim_{N\rightarrow \infty }\frac{f- g}{N^3} =0\;.  
\end{align}
Since the black hole is made of $N$ M5-branes, $S_{BH}$ is expected to scale as $o(N^3)$. 

In this section, we consider the large $N$ limit of the 4d twisted index $\CI_{M_3}(\CT_N[\Sigma_g])$ using the relation in \eqref{4d twisted index from torsions}. For $g > 1$, indeed the twisted index \eqref{4d twisted index from torsions} behaves as $N^3$ at large $N$ limit. Furthermore, we will show that the leading term of $\log \mathcal{I}_{M_3}(\CT_{N}[\Sigma_g])$ perfectly agree with the Bekenstein-Hawking entropy of the magnetically charged black hole in $AdS_5$.

\subsection{Full perturbative $1/N$ expansion of the index}
The $1/N$ pertubative expansion of the 3d twisted index $ \CI_{\Sigma_g}(\CT_N[M_3])$ is studied in \cite{Gang:2018hjd,Gang:2019uay} using the 3d-3d relation. %in \eqref{index from torsion}. 
From the 4d-3d relation \eqref{4d-3d relation},  the 4d twisted index $\CI_{\Sigma_g}(\CT_N[M_3])$ is expected to share the same $1/N$ expansion. Let us summarize the process of taking the large $N$ limit to \eqref{index from torsion}, in order to figure out leading large $N$ terms of the $\CI_{\Sigma_g}(\CT_N[M_3])$.

In the large $N$ limit, only two irreducible flat connections, $\mathcal{A}^{\rm geom}$ and $\mathcal{A}^{\rm \overline{geom}}$ are expected to give exponentially dominant contribution to the summation in \eqref{4d twisted index from torsions}.
\begin{align}
\begin{split}
\mathcal{I}_{M_3} (T_{N}[\Sigma_g]) \xrightarrow{\qquad N\rightarrow \infty \quad } &\; (N \times  {\bf Tor}^{ \textrm{geom}}_{M_3} )^{g-1} +  (N \times  {\bf Tor}^{ \overline{\rm geom}}_{M_3} )^{g-1}   
\\
&+ \textrm{(exponentially smaller terms at large $N$)}\;.
\end{split}
\end{align}
The two dominant flat-connections can be constructed from the unique unit hyperbolic structure on $M_3$:
\begin{align}
\mathcal{A}^{\rm geom}_{N} := \rho_N (\omega +i e)\;, \quad \mathcal{A}^{\overline{\rm geom} } := \rho_N (\omega- i e)\;.
\end{align}
Here $\omega$ and $e$ are spin-connection and dreibein of the hyperbolic structure on $M_3$. 
Both of them can be thought as $so(3)$-valued 1-form on $M_3$ and they form two $PGL(2,\mathbb{C})= SO(3)_{\mathbb{C}}$ flat connections $\mathcal{A}^{\textrm{geom},\overline{\rm geom}}_{N=2} =  \omega \pm  i e$ related to each other by complex conjugation. 
$\rho_N$ is the principal embedding from $PGL(2,\mathbb{C})$ to   $PGL(N, \mathbb{C})$.

Using mathematical results \cite{muller2012asymptotics,park2019reidemeister}, we obtain  following  asymptotic expansion of the twisted index at large $N$ \cite{Gang:2018hjd,Gang:2019uay}
\begin{align}
\begin{split}
\mathcal{I}_{M_3} (T_{N}[\Sigma_g])   & = \left(e^{i \theta(N,M_3)} +e^{-i \theta (N,M_3)}\right) \;
\\
& \ \times  \exp \bigg{[}(g-1) \left( \frac{\textrm{vol}(M_3)}{6\pi } (2N^3-N-1) + \log N\right) \bigg{]} \;
\\
& \ \times  \exp \bigg{[}(g-1) \left( \sum_{\gamma}\sum_{m=1}^{N-1} \sum_{k=m+1}^{\infty}  \log |1-e^{-k \ell_{\mathbb C}(\gamma)} | \right) \bigg{]} \; 
\\
&+ (\textrm{exponentially smaller terms at large $N$})  \;. \label{large N expansion}
\end{split}
\end{align}
The exponentially smaller terms come from  contributions of irreducible flat connections other than $\mathcal{A}^{\rm geom}_{N}$ and $\mathcal{A}^{\overline{\rm geom}}_{N}$. Again, the above expansion only  holds for closed hyperbolic 3-manifold with vanishing $H_1(M_3, \mathbb{Z}_N)$. $\theta (N,M_3)$ is an undetermined angle due to the relative phase differences of the contributions from two dominant flat-connections, $\mathcal{A}^{\rm geom}$ and $\mathcal{A}^{\rm \overline{geom}}$. $\sum_{\gamma}$ is  summation over the nontrivial primitive conjugacy classes  of  the fundamental group $\pi_1(M_3)$. $\ell_{\mathbb{C}}(\gamma)$ is the complexified geodesic length of the $\gamma$, which is defined by following relation:
\begin{align}
\textrm{Tr} P\exp \left(-\oint_{\gamma} \CA_{N=2}^{\rm geom} \right)= e^{\frac{1}2 \ell_{\mathbb{C}}(\gamma)}+ e^{-\frac{1}2 \ell_{\mathbb{C}}(\gamma)}\;, \quad \mathfrak{Re} [\ell_{\mathbb{C}}]>0\;.
\end{align}
The term $\Sigma_\gamma (\ldots)$ in the above  can be decomposed into two parts
\begin{align}
\begin{split}
&\sum_{\gamma}\sum_{m=1}^{N-1} \sum_{k=m+1}^{\infty}  \log |1-e^{-k \ell_{\mathbb C}(\gamma)} | 
\\
&= -\mathfrak{Re} \sum_\gamma \sum_{s=1}^{\infty}\frac{1}s \left(\frac{e^{-s \ell_{\mathbb{C}}}}{1-e^{-s  \ell_{\mathbb{C}}}}\right)^2 + \mathfrak{Re} \sum_\gamma \sum_{s=1}^{\infty}\frac{1}s \left(\frac{e^{-\frac{s(N+1)}{2} \ell_{\mathbb{C}}}}{1-e^{-s  \ell_{\mathbb{C}}}}\right)^2
\end{split}
\end{align} 
The first term is $N$-independent while the second term is exponentially suppressed at large $N$. Note that the leading order behavior of the twisted index only depend on vol$(M_3)$ while subleadings depend on both vol$(M_3)$ and length spectrum $\{\ell_{\mathbb{C}}(\gamma) \}$.

\subsection{A magnetically charged $AdS_5$ black hole }
In \cite{Nieder:2000kc}, a magnetically charged black hole solution in 5d $\mathcal{N}=4$  gauged supergravity is numerically constructed.  The black hole interpolates
\begin{align}
\begin{split}
&\textrm{UV  :  $AdS_5$ with $\mathbb{R}\times M_3$ conformal boundary}
\\
&\textrm{IR : $AdS_2 \times M_3$ near-horizon}
\end{split}
\end{align}
The 11d uplift of the solution is studied in \cite{Gauntlett:2007sm}. Including the internal direction, the near horizon geometry is a warped product of $AdS_2 \times M_3 \times \Sigma_g \times S^4$ \cite{Gauntlett:2001jj}. The Bekenstein-Hawking entropy of the $AdS_5$ black hole is universally given as \cite{Bobev:2017uzs}
\begin{align}
S_{BH} = a_{4d}\frac{\textrm{vol}(M_3)}{\pi}\;.
\end{align} 
Here $a_{4d}$ is the a-anomaly coefficient of the dual 4d $\mathcal{N}=2$ SCFT. For class $\CS$ theory $\CT_{N}[\Sigma_g]$, the anomaly coefficient is given by \cite{Gaiotto:2009gz}
\begin{align}
a_{4d}(\CT_{N}[\Sigma_g]) = \frac{1}3 (g-1) N^3 + \textrm{subleading},
\end{align}
thus the Bekenstein-Hawking  entropy for the magnetically charged black hole inside $AdS_5$ dual \cite{Maldacena:2000mw,Gaiotto:2009gz} of 4d $\CT_{N}[\Sigma_g]$ read
\begin{align}
S_{BH} = \frac{(g-1) N^3 \textrm{vol}(M_3)}{3\pi}\;. \label{S-BH}
\end{align}
Here $\textrm{vol}(M_3)$ is the volume of the 3-manifold measured using unit hyperbolic metric normalized as
\begin{align}
R_{\mu\nu} = -2 g_{\mu\nu}\;.
\end{align}
Locally, the unit hyperbolic metric is given as
\begin{align}
ds^2 (\mathbb{H}^3) = d\phi^2+\sinh^2 \phi (d\theta^2+ \sin^2 \theta d\nu^2)\;.
\end{align}
According to Mostow's rigidity theorem \cite{mostow1968quasi}, there is an unique unit hyperbolic metric on $M_3$ and the volume is actually a topological invariant.

It turns out that the large $N$ leading part of the $\log \mathcal{I}_{M_3}(\CT_{N}[\Sigma_g])$  nicely matches with the Bekenstein-Hawking entropy \eqref{S-BH}:
\begin{align}
\log \mathcal{I}_{M_3}(\CT_{N}[\Sigma_g]) = \frac{(g-1) N^3 \textrm{vol}(M_3)}{3\pi} + (\textrm{subleadings})
\end{align}  
From \eqref{large N expansion}, the full perturbative  $1/N$ corrections can be summarized as follows
\begin{align}
\begin{split}
&\textrm{a term with $\textrm{vol}(M_3)$ \; :\;  }(g-1)  \frac{\textrm{vol}(M_3)}{6\pi } (2N^3-N-1)\;,
\\
&\textrm{a term involving length spectrum : } (1-g)\times \mathfrak{Re} \sum_\gamma \sum_{s=1}^{\infty}\frac{1}s \left(\frac{e^{-s \ell_{\mathbb{C}}}}{1-e^{-s  \ell_{\mathbb{C}}}}\right)^2 \;,
\\
&\textrm{logarithmic correction : } (g-1)\log N\;. \label{perturbative corrections}
\end{split}
\end{align}
Since we do not have understanding of contributions from other irreducible flat-connections, it is very difficult to determine the full non-perturbative corrections. From the above analysis, we could  identify  following non-perturbative correction:
\begin{align}
\textrm{a non-perturbative correction : }(g-1)\mathfrak{Re} \sum_\gamma \sum_{s=1}^{\infty}\frac{1}s \left(\frac{e^{-\frac{s(N+1)}{2} \ell_{\mathbb{C}}}}{1-e^{-s  \ell_{\mathbb{C}}}}\right)^2\;. \label{non-perturbative  correction}
\end{align}
 It would interesting future work to identify these subleading corrections in \eqref{perturbative corrections} and \eqref{non-perturbative  correction} from the quantum computation in the holographic dual M-theory. Especially, the logarithmic subleading correction should be reproduced from zero mode analysis on the 11d uplift of the $AdS_5$ black hole as done in \cite{Bhattacharyya:2012ye,Liu:2017vbl,Gang:2019uay} for AdS$_4$/CFT$_3$ examples.
 
 %%%%%%%%%%%%%%%%%%%%%%%%%%%%%%%%%%%%%%%%%%%%%%%%%%%%%%%%
 \section*{Acknowledgments}
%This work was partially done while one of  authors (DG) was visiting APCTP, Pohang for a workshop ``Frontiers of Physics Symposium'', 13-14 May 2019. We thank APCTP for hospitality.  The research of DG was partially supported by NRF grant 2019R1A2C2004880.
This work was partially done while one of  authors (DG) was visiting APCTP, Pohang for a workshop ``Frontiers of Physics Symposium'', 13-14 May 2019. We thank APCTP for hospitality.  The researches of DG and KL were supported in part by the National Research Foundation of Korea Grant  NRF grant 2019R1A2C2004880 and NRF-2017R1D1A1B06034369, respectively. This work was benefited from the 2019 Pollica summer workshop, which was supported in part by the Simons Foundation (Simons Collaboration on the Non-perturbative Bootstrap) and in part by the INFN.

\appendix
%\section{Derivation of the univeral relation $S_{BH} = a_{4d} \frac{\textrm{vol}(M)}{\pi}$ from background field analysis}

\bibliographystyle{ytphys}
\bibliography{ref-AdS5}

\end{document}